\documentclass[pdflatex,sn-nature]{sn-jnl}

\usepackage[normalem]{ulem}
\usepackage{graphicx}%
\usepackage{multirow}%
\usepackage{amsmath,amssymb,amsfonts}%
\usepackage{amsthm}%
\usepackage{mathrsfs}%
\usepackage[title]{appendix}%
\usepackage[dvipsnames]{xcolor}%
\usepackage{textcomp}%
\usepackage{manyfoot}%
\usepackage{booktabs}%
\usepackage{algorithm}%
\usepackage{algorithmicx}%
\usepackage{algpseudocode}%
\usepackage{listings}%
\usepackage{setspace}
\usepackage{physics}
\usepackage{ulem}

\newcommand*{\ab}[1]{\textcolor{black}{#1}}
\newcommand*{\mj}[1]{\textcolor{black}{#1}}

\theoremstyle{thmstyleone}%
\theoremstyle{thmstyletwo}%

\theoremstyle{thmstylethree}%

\begin{document}

\title[Article Title]{Persistent singlet electronic character in the multiexcitonic triplet-pair state of strongly coupled pentacene singlet fission dimers}



\author[1]{\fnm{Atandrita} \sur{Bhattacharyya}}
\author[2]{\fnm{Namana} \sur{Venkatareddy}}
\equalcont{These authors contributed equally to this work.}
\author[1]{\fnm{Sanjoy} \sur{Patra}}
\equalcont{These authors contributed equally to this work.}
\author[1]{\fnm{Kanad} \sur{Majumder}}
\equalcont{These authors contributed equally to this work.}
\author[1]{\fnm{Vithoba} \sur{Hugar}}

\author*[1]{\fnm{Satish} \sur{Patil}}\email{spatil@iisc.ac.in}
\author*[2]{\fnm{Manish} \sur{Jain}}\email{mjain@iisc.ac.in}
\author*[1]{\fnm{Vivek} \sur{Tiwari}}\email{vivektiwari@iisc.ac.in}


\affil*[2]{\orgdiv{Department of Physics}, \orgname{Indian Institute of Science}, \orgaddress{\street{CV Raman Road}, \city{Bengaluru}, \postcode{560012}, \state{Karnataka}, \country{India}}}
\affil*[1]{\orgdiv{Solid State and Structural Chemistry Unit}, \orgname{Indian Institute of Science}, \orgaddress{\street{CV Raman Road}, \city{Bengaluru}, \postcode{560012}, \state{Karnataka}, \country{India}}}


\maketitle

\begin{abstract}

Singlet fission converts an optically excited singlet state into a spin-entangled triplet pair state $\mathrm{(TT_1)^1}$ that can, in principle, yield two free triplets for photovoltaics and/or a polarized high spin state for quantum technologies. Synthetically tunable templates suggest that the above photophysics is governed by a subtle but poorly understood interplay of molecular motifs, geometry and structural fluctuations. Here, we investigate the $\mathrm{(TT_1)^1}$ state in a library of conformationally flexible pentacenic dimers, where a $\mathrm{(TT_1)^1}$-specific near-IR spectral feature is readily available. Using a suite of polarization-controlled impulsive optical spectroscopies, we find that $\mathrm{(TT_1)^1}$ formation is specific to planar conformations and is accompanied by large nuclear reorganization in the $\mathrm{(TT_1)^1}$ photoproduct. Introducing polarization anisotropy to {track the electronic character of the $\mathrm{(TT_1)^1}$ species}, supported by {screened configuration interaction based electronic structure theory}, we find that significant singlet-triplet electronic mixing is persistent throughout its evolution. This behavior is universal across diverse bridging motifs and indicates that, once the triplet pair is strongly bound, neither substantial nuclear reorganization nor structural fluctuations on longer timescales are sufficient to suppress persistent singlet--triplet electronic mixing, such that triplet-pair decorrelation is outcompeted by its decay. Our observations establish polarization-selective pump--probe and anisotropy as a direct optical probe of triplet pair decorrelation, complementary to spin-selective measurements at longer timescales.

\end{abstract}

\section*{Introduction}
Singlet exciton fission\cite{MichlARPC2013} (SF) is an ultrafast internal conversion process, where an optically excited singlet state converts into a spin-entangled triplet pair intermediate with an overall singlet spin, denoted as $\mathrm{(TT_1)^1}$, on timescales as fast as sub-100 fs. At longer timescales, the triplet pair intermediate can yield a long-lived, polarized high-spin state at elevated temperatures\cite{Smyser2020Singlet, Dill2023Entangled} with potential applications in quantum information and hyperpolarization NMR\cite{Yanai2023}, and/or go on to dissociate into two free triplets with potential photovoltaic applications. Synthetically tunable templates for intramolecular SF\cite{Johnson2020} (iSF) provide a useful model system from which the effect of molecular design parameters on $\mathrm{(TT_1)^1}$ dynamics can be systematically understood. While the proposed mechanisms for $\mathrm{(TT_1)^1}$ formation have ranged from direct\cite{MichlARPC2013}, superexchange\cite{Reichman2014} and charge transfer (CT) mediated\cite{ZhuARPC2015} electronic couplings, and vibrational-electronic couplings\cite{Damrauer2010,DamrauerJPCA2014,RaoNatChem2016, CamposACSCentSci2016, Herbert2017,TempelaarJCP2018,Duan2020,Bhattacharyya2023Low}, the subtle interplay of molecular motifs\cite{BardeenJACS2007,GuldiPNAS2015,Campos2017,BradforthJACS2018,PatilJACS2023,PatilAngew2024,PatilCellReports2024,Bansal2022highly}, relative geometry\cite{BardeenJACS2007,BradforthJACS2018,Kay2018_NatComm,Hasobe2018_JPCL,DamrauerJACS2019,ThossJChemPhys2019} and structural fluctuations\cite{Wasielewski2019_PNAS_QuintetTripletMixing,Kim2024_JACS,Hasobe2020_JPCB} that governs the efficient formation of $\mathrm{(TT_1)^1}$ and its subsequent spin-selective conversion into desirable photophysical channels constitutes the central unresolved mechanistic question in singlet fission. \par

Since the $\mathrm{T_1 \rightarrow T_n}$ transition dipole in the photoproduct is orthogonal\cite{Kohn2008,Schreiber2012,Shukla2014} to the S$_0 \rightarrow$ S$_1$ and S$_1 \rightarrow$ S$_n$ transition dipoles\cite{Friend_PRB_2011}, significant electronic reorientation during singlet fission is expected. However, even though ultrafast $\mathrm{(TT_1)^1}$ formation and evolution has been widely reported, the associated electronic reorientation remains largely unexplored. Spin evolution on longer timescales is {routinely probed\cite{Hasobe2018_JPCL,Hasobe2020_JPCB,Kobori2020_ChemSci,Tayebjee2016_NatPhys,Wasielewski2019_PNAS_QuintetTripletMixing} by time-resolved EPR to infer decorrelation of the $\mathrm{(TT_1)^1}$ state into quintets and triplets, $\mathrm{(TT_1)^5}$ and $\mathrm{(TT_1)^3}$, respectively}. This work demonstrates that tracking the electronic character during S$_1 \rightarrow$ $\mathrm{(TT_1)^1}$ conversion and its subsequent evolution into $\mathrm{(TT_1)^m}$ and free triplets, where the superscript $\text{m}$ denotes the spin manifold, can provide key mechanistic insights about $\mathrm{(TT_1)^1}$ decorrelation, {complementary to spin-selective measurements at longer timescales.} \par

While structurally well-defined dimers have recently\cite{Dill2023Entangled} demonstrated nearly perfectly polarized generation of $\mathrm{(TT_1)^5_{m_s=0}}$ quintet spin sub-level at 77K, structural fluctuations in conformationally flexible iSF systems\cite{GuldiEurJourChem2018,PatilJACS2023} are now believed\cite{BradforthJACS2018,Wasielewski2019_PNAS_QuintetTripletMixing,Kim2024_JACS,Hasobe2020_JPCB} to play an important functional role as well -- they can fine-tune $\mathrm{(TT_1)^1}$--$\mathrm{(TT_1)^5}$ and $\mathrm{(TT_1)^5}$--$\mathrm{(TT_1)^3}$ mixing pathways through terahertz (THz) vibrational motions\cite{Hasobe2020_JPCB,Kobori2020_ChemSci} thereby enabling decorrelation of the spin manifold over undesirable TT decay via internal conversion. One complication in the conformationally flexible iSF systems is the presence of multiple ground state conformations which can lead to a heterogeneous distribution of through bond and through space couplings\cite{BardeenJACS2007}. Consequently, heterogeneous iSF rates\cite{GuldiJPCLett2022,PatilCellReports2024,PatilJACS2023} {are to be expected}. Apart from overlapping spectral lineshapes from different conformations, yet another layer of complexity in tracking the time-evolving electronic character of the $\mathrm{(TT_1)^1}$ state in model iSF dimers arises from the broad and overlapping spectral signatures\cite{Scholes2016,MazumdarJPCLett2017,Mazumdar2018,Mazumdar2020} of the $\mathrm{(TT_1)^1}$ state compared to those of the free triplets. Only recently has a distinct near-IR signature of the $\mathrm{(TT_1)^1}$ state was reported\cite{CamposACSCentSci2016} in pentacenic (Pc) iSF dimers, notably\cite{XYZ2017} with chemical reactivity different from free triplets. Extensive theoretical investigation\cite{MazumdarJPCLett2017,Mazumdar2018,Mazumdar2020} has suggested that the near-IR spectral feature is unique to the bound triplet state $\mathrm{(TT_1)^1}$ with its strength expected to increase with electronic coupling. For example, the strength of the near-IR excited state absorption (ESA) band increases with crystalline packing\cite{Scholes2018}. \par

Here using polarization-controlled two-dimensional electronic spectroscopy (2DES) we selectively probe the electronic character of the $\mathrm{(TT_1)^1}$-specific near-IR spectral feature in conformationally flexible Pc dimers across a variety of bridging units. The additional excitation axis in the 2D maps, spanning $\sim$300 nm, directly reports on the near-planar conformations that lead to the rapid $\mathrm{(TT_1)^1}$ formation and are associated with a distinct near-IR ESA band. The evolution of $\mathrm{(TT_1)^1}$ is accompanied by enhanced vibrational quantum-beat amplitudes in the $\mathrm{(TT_1)^1}$ photoproduct, spanning intramolecular vibrational modes from $\sim$250--1200~cm$^{-1}$ and indicating substantial nuclear reorganization during this process. However, polarization-controlled impulsive pump-probe experiments, which selectively probe\cite{ZanniPNAS2022,Bhattacharyya2023Low,sahu2025isolating} perpendicularly polarized initial and final transitions---for example, S$_0 \rightarrow$ S$_1$ and $\mathrm{(TT_1)^m \rightarrow (TT_n)^m}$--directly reveal that the $\mathrm{(TT_1)^1}$ state retains a significant degree of singlet electronic character.  Polarization anisotropy experiments time-resolve the electronic character of $\mathrm{(TT_1)^1}$ and further corroborate this observation -- the singlet electronic character in the $\mathrm{(TT_1)^1}$ state persists throughout its evolution away from the Franck-Condon (FC) region until its eventual decay. \mj{We further support our findings with screened configuration interaction based many-body electronic structure theory\cite{Jain2025} which reproduce the intermediate polarization of the near-IR ESA band, in qualitative agreement with the experiments.} Our observations are {found to be general} across a variety of bridging motifs and therefore carry important implications with regard to the role of structural fluctuations in the decorrelation of the $\mathrm{(TT_1)^1}$ state. Evidently, {once the triplet pair is strongly bound to give rise to a prominent near-IR ESA band}, the substantial nuclear reorganization accompanying $\mathrm{(TT_1)^1}$ evolution is {unable} to suppress the undesirable singlet-triplet electronic mixing in the $\mathrm{(TT_1)^1}$ state or effectively compete against the internal conversion channels for $\mathrm{(TT_1)^1}$ decay. {These observations provide an important design consideration for iSF systems targeting long-lived high-spin states or efficient triplet separation into free triplets; a larger statistical space that allows for rapid triplet diffusion\cite{TempelaarJCP2018,Kim2024_JACS,Johnson2020_NatChem,Kobori2018} to neighboring chromophores may be key for decorrelating the strongly bound triplet pair.} \par


\newpage

\section*{Results and discussion}

\subsection*{Steady state measurements}  
\begin{figure*}[!ht]
	\centering
	\includegraphics[width=5 in]{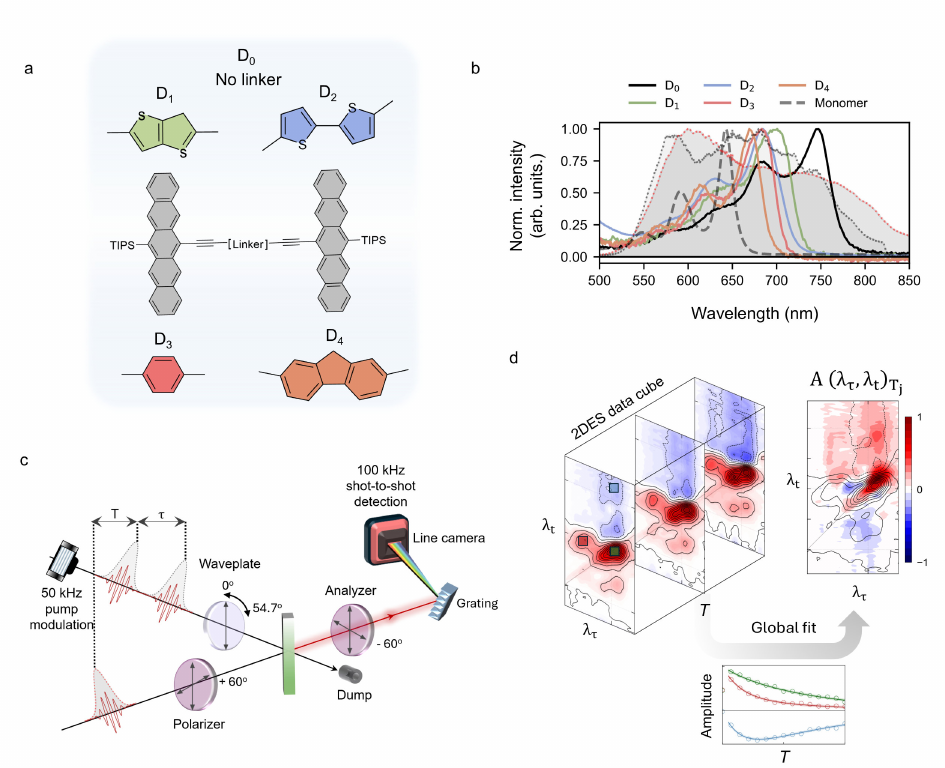} 
	\caption{ \footnotesize \textbf{Pentacene (Pc) dimers with varying linkers and polarization-controlled two-dimensional electronic spectroscopy (2DES).} \textbf{a} Molecular structure of the Pc dimers $\mathrm{D_0}$--$\mathrm{D_4}$ with the different linkers. \textbf{b} Normalized steady-state absorption spectra of covalently linked Pc dimers $\mathrm{D_0}$--$\mathrm{D_4}$ in chlorobenzene and the Pc monomer (black dashed trace) in toluene solution. The pump and the probe spectra are shown as shaded gray area with black and red dotted lines, respectively. \textbf{c} Schematic illustration of 100 kHz shot-to-shot 2DES spectrometer\cite{Thomas2023Rapid,Bhat2023Rapid} in partially collinear pump-probe geometry. The white light continuum (WLC) pump and probe spectra are generated from a YAG nonlinear crystal. Identical replicas of the pump pulse are generated using a common-path interferometer which creates the delay $\tau$ between the pump pulses. The probe enters the sample after a pump-probe waiting time delay $T$. The absorptive changes in the sample are detected using a 100 kHz shot-to-shot CCD line camera detection scheme\cite{Thomas2023Rapid,Bhat2023Rapid}. \textbf{d} 2DES experiment collects 2D maps that correlate detection with excitation frequencies (or wavelengths), as a function of $T$. Global fitting of the population background leads to rate maps (RMs) which are 2D analogues of decay associated spectra (DAS).}
	\label{fig:fig1}
\end{figure*}
{Majumder et al. have} recently reported Pc dimers connected at 6,6' {and 2,2'} positions with acetylene groups spaced by a series of linkers\cite{PatilJACS2023,PatilAngew2024}. Using non-impulsive pump-probe experiments, ps to sub-ps iSF and $\mathrm{(TT_1)^1}$ dynamics could be characterized across a variety of planar and twisted bridging motifs. In this study, we focus on the Pc dimers which exhibit the near-IR ESA feature that contributes as a $\mathrm{(TT_1)^1 \rightarrow (TT_n)^1}$ absorption and was previously attributed to be specific to the coupled triplet biexcitons\cite{CamposACSCentSci2016, XYZ2017, MazumdarJPCLett2017}, that is, the $\mathrm{(TT_1)^1}$ state. In the chosen 6,6'-linked dimers the linker spans five different moieties (no spacer ($\mathrm{D_0}$), thienothiophene ($\mathrm{D_1}$), bithiophene ($\mathrm{D_2}$), phenyl ($\mathrm{D_3}$), fluorene ($\mathrm{D_4}$)), as shown in {Fig.~\ref{fig:fig1}a}. In the five different linker bridges, the $\mathrm{(TT_1)^1}$ formation rates were reported to be progressively slower. All the dimers show planar optimized geometries, with the exception of $\mathrm{D_2}$ with a dihedral angle of $\sim$15$^\circ$, although heterogeneous conformations should be expected at room temperature. The steady state absorption spectra of the Pc dimers in chlorobenzene solution are shown in {Fig.~\ref{fig:fig1}b}. The TIPS-Pc monomer spectrum measured in toluene is also overlaid. Prominent intensity bands, akin to 0-0 and 0-1 Franck-Condon (FC) vibrational progressions, can be observed in all cases. The filled light and dark gray traces denote the pump and the probe laser spectra. The dimers show significant red-shift compared to the monomer -- 0-0 peak is situated at 750~nm for $\mathrm{D_0}$, 700~nm for $\mathrm{D_1}$, 690~nm for $\mathrm{D_2}$, 685~nm for $\mathrm{D_3}$, 675~nm for $\mathrm{D_4}$, versus 642~nm for the monomer. A notable broadening of the main band also correlates well with the observed red-shift. The absorption spectra upon $\sim$100$\times$ dilution are shown in the {Supplementary Fig.~S4} and do not show any shifts, thus ruling out aggregation. \par

Fig.~\ref{fig:fig1}c shows the schematic of the polarization-controlled two-dimensional electronic spectroscopy (2DES) and pump-probe setup\cite{Bhat2023Rapid,Thomas2023Rapid,ThomasNatComm2026} (Supplementary Note 1) with independent polarization control in the pump, probe and detection paths. The pump and probe white light inputs are generated by a YAG nonlinear crystal, resulting in spectral bandwidth of $>$300 nm (Fig.~\ref{fig:fig1}b). As shown in Fig.~\ref{fig:fig1}d, 2DES creates a correlation map of detection versus excitation frequency of the system at different pump-probe waiting times $T$. Heterogeneous iSF rates can be expected \cite{BardeenJACS2007, BradforthJACS2018} due to modest rotational barriers around the acetylene linkage. Unlike the narrow band pump-probe experiments, the excitation frequency axis $\mathrm{\omega_{\tau}}$ in 2DES is created by Fourier transformation of the pump pulse pair delay $\mathrm{\tau}$. This preserves the temporal resolution necessary for resolving fast $\mathrm{(TT_1)^1}$ formation, while simultaneously providing spectral resolution along the excitation axis necessary to spectrally decongest the overlapping spectral bands from heterogeneous dimer conformations. As we will demonstrate below in Fig.~\ref{fig:fig2}, a further degree of spectral decongestion along the excitation axis becomes possible when the population decay in the 2DES data cube is resolved as 2D rate maps (RMs). RM for a given time constant $\mathrm{T_j}$, denoted as $\mathrm{A(\lambda_{\tau},\lambda_t)_{T_j}}$,  is the 2D analogue of the decay associated spectrum (DAS). Conformations which lead to rapid $\mathrm{(TT_1)^1}$ formation, seen as growth of the near-IR band, can be identified based on the rise/decay amplitude along the excitation wavelength axis $\mathrm{\lambda_{\tau}}$, obtained by Jacobian transformation of the frequency axis $\mathrm{\omega_{\tau}}$. \par 



\subsection*{2D rate maps reveal conformation specific $\mathrm{(TT_1)^1}$ formation}

\begin{figure*}[h]
	\centering
	\includegraphics[width=4.5 in]{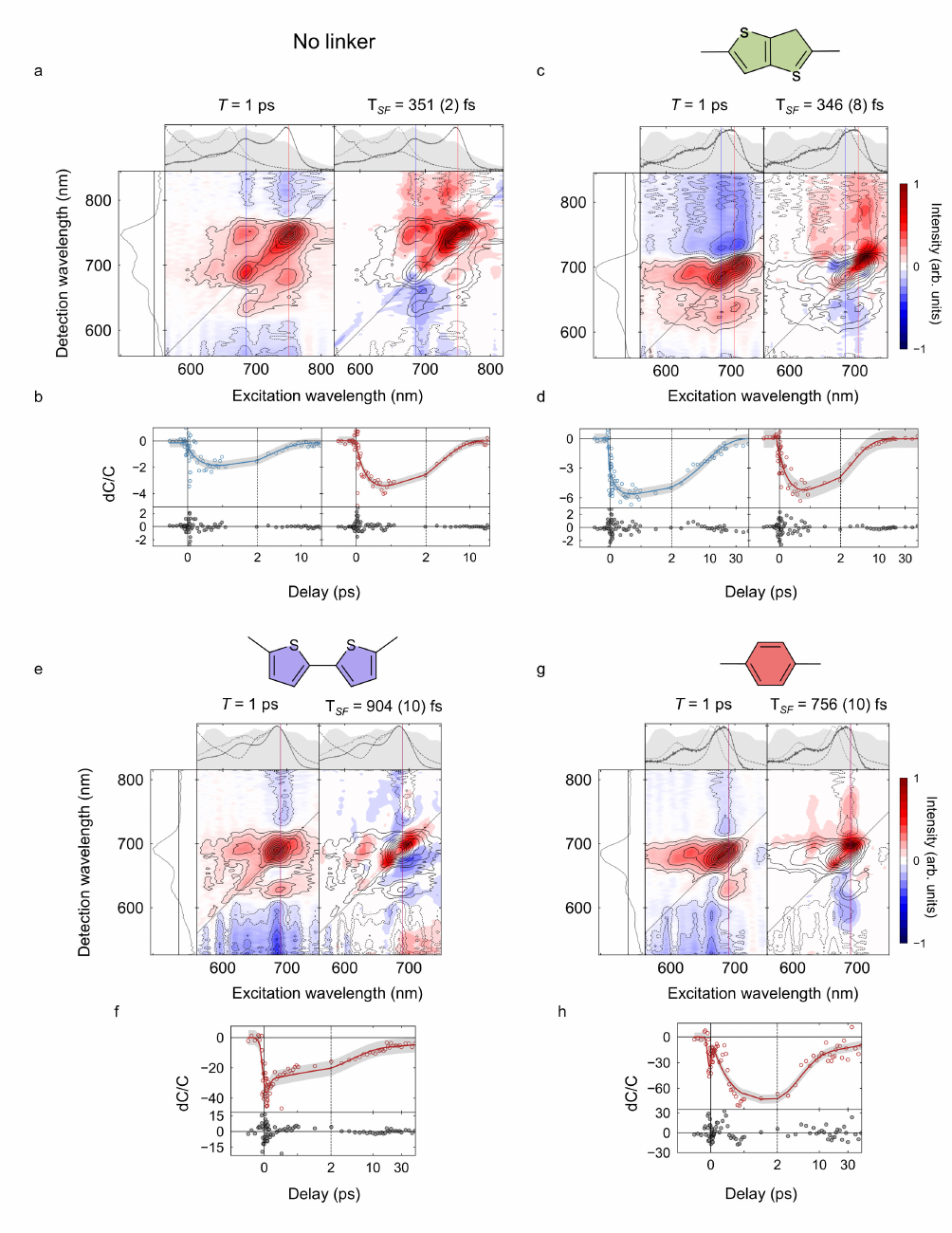}
	\caption{\footnotesize \textbf{Conformation specific and excitation wavelength independent $\mathrm{(TT_1)^1}$ formation revealed by 2DES.} Normalized absorptive 2D spectrum at $T$ = 1 ps for samples $\mathrm{D_{0-3}}$ is shown in the left panel of \textbf{(a)}, \textbf{(c)}, \textbf{(e)} and \textbf{(g)} respectively. The top panels overlay the absorption (ABS, solid), emission (EMS, dashed) and fluorescence excitation (FLEX, dotted) spectra. The filled gray shaded area denotes the pump spectrum. The vertical inset panel on the left shows the spectrally integrated 2D spectrum at the same $T$ time delay. The corresponding right 2D panels show the kinetic RMs for the $\mathrm{(TT_1)^1}$ formation time constant denoted by T$_{SF}$. The contour lines in RMs show the absorptive 2DES at 1~ps. The contour lines are spaced at 5\%, and 10-100\% in 10\% steps. The 2DES data cube is collected from a stepwise $T$ rate scan with 15-20 points per decade. The RM is obtained by fitting the 2DES to a global model obtained by a corresponding spectrally integrated 2DES dataset. The fitting details are presented in {Supplementary Note S3.3}. $\mathrm{D_{4}}$ is not shown because its weaker NIR ESA band is not prominently seen in the lower SNR rate scans (Supplementary Fig.~S22). The early $T$ 2D spectra of $\mathrm{D_{4}}$ obtained from coherence scans and shown in Supplementary Fig.~S24 clearly show the NIR band. The red and blue vertical lines denote excitation wavelength slices, possible for $\mathrm{D_{0-1}}$. For $\mathrm{D_2, D_3}$, no clear red and blue excitation bands are available in the near-IR region. Hence the excitation band is integrated along $\lambda_{\tau}$ axis. Panels \textbf{(b, d, f, h)} show the near-IR ESA signal integrated over $\sim$ 70-90~nm near-IR $\lambda_t$ band for a given excitation slice. The color-coded solid lines denote the fits with the $\pm \sigma$ standard deviation shown as a error band. The error band is calculated from multiple 2D spectra at $T$ = 0.5 ps interleaved across data collection. The bottom panels denote the corresponding residuals. For $\mathrm{D_{0-1}}$, $\mathrm{(TT_1)^1}$ formation is found to be excitation wavelength independent with error. However, for $\mathrm{D_{1}}$ the $\mathrm{(TT_1)^1}$ decay timescale is $\sim$3x slower for blue excitation wavelength (Supplementary Fig.~S15), well beyond experimental and fitting errors. The fitting details are tabulated in {Supplementary Table~S6-S8}. }
	\label{fig:fig2}
\end{figure*}

Fig.~\ref{fig:fig2} presents 2D absorptive spectra at $T$ = 1 ps (left in each panel a,c,e,g) for dimers $\mathrm{D}_{0-3}$. The top inset on each 2D spectrum overlays the absorption (ABS), emission (EMS) and fluorescence-excitation (FLEX) spectra for each case. Compared to a Stokes-shifted EMS expected in a monomer, the dimer EMS in diluted solutions, to suppress any fluorescence reabsorption effects, peak at substantially bluer wavelengths. $\mathrm{D_0}$ shows the maximum shift of $\sim$80 nm which is consistent with an earlier report \cite{GuldiJPCLett2022} on $\mathrm{D_0}$ dimer where planar conformations were found to be red-shifted and the ones that participate in ultrafast $\mathrm{(TT_1)^1}$ formation. The emissive conformations are blue-shifted and clearly identified by the overlaid FLEX spectra in Fig.~\ref{fig:fig2}. Comparison of the FLEX and ABS reveals that the observed broadening within the I$_{00}$ intensity band in the absorption spectra (Fig.~\ref{fig:fig1}b) partly also arises due to the blue-shifted emissive conformations of the Pc dimer. 

In the corresponding 2D spectra, the prominent 0-0 and 0-1 like vibronic bands in the ABS contribute as diagonal (DP) and cross peaks (CP). The ESA bands are seen on red and blue sides of the positive bands. The blue ESA band has been previously assigned to T$_1$ $\rightarrow$ T$_n$ transitions in Pc thin films\cite{Lochbrunner_PRL_2007, Friend_PRB_2011}. The near-IR ESA band which is specific to $\mathrm{(TT_1)^1}$ is seen in all cases, however with intensity concentrated at the redder excitation wavelengths corresponding to the non-emissive lowest energy absorption peak overlaid in the top panel. Representative 2D spectra for all the dimers are also shown in Supplementary Figs.~S14-S22. Among these dimers, even though $\mathrm{D_2}$ is not the most strongly coupled\cite{PatilJACS2023}, in terms of reduced red-shift in ABS and iSF rates (Supplementary Table S5-S6), a prominent NIR band is seen in the 2D spectrum ({Supplementary Fig.~S24}) even at the earliest $T$ within the instrument response. Similar but lesser signal-to-noise ratio (SNR) trends are also seen for $\mathrm{D_4}$. This suggests direct photoexcitation of the $\mathrm{(TT_1)^1}$ state. We have further confirmed these features with higher SNR pump-probe measurements discussed in Fig.~\ref{fig:fig5} and Supplementary Fig.~31. \par

For kinetic analysis of the 2D dataset, we first ignore excitation wavelength dependent $\mathrm{(TT_1)^1}$ dynamics, if any. The corresponding RM amplitude for the fastest time constant T$_{SF}$ is shown in Fig.~\ref{fig:fig2} panels b,d,f,h for the dimers $\mathrm{D_{0-3}}$, respectively. The fitting details are presented in {Supplementary Note S3.3}. Note that the 2D RMs correspond to the fastest time constants shown on the top of each panel. Red denotes positive amplitude associated with a given time constant. A positive amplitude in the negative near-IR ESA region implies a rising amplitude which corresponds to the formation of $\mathrm{(TT_1)^1}$ species. The 2D RMs show a distinct concentration of amplitude towards the red-most wavelengths, even more so than that reflected in the 2D spectra. This amplitude distribution aligns well with the planar, red-shifted and non-emissive conformations indicating that fast $\mathrm{(TT_1)^1}$ formation is conformation specific. This provides direct experimental evidence supporting the theoretical prediction\cite{Mazumdar2018} that the bound $\mathrm{(TT_1)^1}$ state, characterized by entangled triplets and its associated near-IR ESA band, is the strongest in planar conformations where electron and hole overlap integrals are maximized. \par 

Our recent 2DES experiments\cite{PatraJCP2026} on a strongly coupled iSF dimer have reported excitation wavelength dependence of $\mathrm{(TT_1)^1}$ formation dynamics. For the Pc dimers, conformation-specific $\mathrm{(TT_1)^1}$ formation revealed by 2D rate maps also allows us to {probe} whether $\mathrm{(TT_1)^1}$ formation and its subsequent evolution {in the $\mathrm{D_{0,1}}$ dimers, the two cases with the most prominent near IR ESA band}, is excitation wavelength dependent {within our excitation bandwidth}. When excitation wavelength dependent $\mathrm{(TT_1)^1}$ dynamics is present, fitting a 2D dataset with a global kinetic model leads \cite{PatraJCP2026} to $T$ dependent as well as $\mathrm{\lambda_{\tau}}$ dependent residual trends in the 2D residual maps. The residual trends resulting from the 2D global kinetic analysis are analyzed in the Supplementary Figs.~S13--S21.  In the 2D residual maps, no residual amplitude trends are seen during $\mathrm{(TT_1)^1}$ formation timescales or in specific excitation wavelength regions. We therefore conclude that $\mathrm{(TT_1)^1}$ formation is not excitation wavelength dependent within experimental error, and within our pump excitation bandwidth. To further confirm excitation-wavelength dependent dynamics we fit the red and blue excitation wavelength slices in Fig.~\ref{fig:fig2} in the near-IR band individually. For $\mathrm{D_{0}}$, the resulting time constants in the two slices are similar within experimental and fitting errors (Supplementary Table~S7). However, as shown for $\mathrm{D_1}$ in Fig.~\ref{fig:fig2}d, Supplementary Fig.~S15 and Table~S8, the $\mathrm{(TT_1)^1}$ decay timescale is $\sim$3$\times$ slower for the blue slice, significantly beyond experimental and fitting errors. {The observation of excitation wavelength dependent near-IR ESA decay is consistent with our recent\cite{PatilJACS2023} pump-probe study on a larger library of Pc dimers with both planar and twisted bridges where the $\sim$520 nm ESA band was probed.} It is important to note that a similar analysis in the DP region can be made complicated by spectral diffusion and overlapping GSB spectral bands from different conformations ({Supplementary Fig.~S11-S12}). Note also the IRF limited appearance of $\mathrm{(TT_1)^1}$ for $\mathrm{D_{2}}$ in Fig.~\ref{fig:fig2}f, confirmed further in Supplementary Fig.~S24 and Fig.~S31. This is further confirmed in Fig.~\ref{fig:fig5} with higher SNR pump-probe experiments.\par 

\clearpage

\subsection*{Enhanced quantum beats on the $\mathrm{(TT_1)^1}$ photoproduct suggest large nuclear reorganization upon its formation}
\begin{figure*}[ht!]
	\centering
	\includegraphics[width=\textwidth]{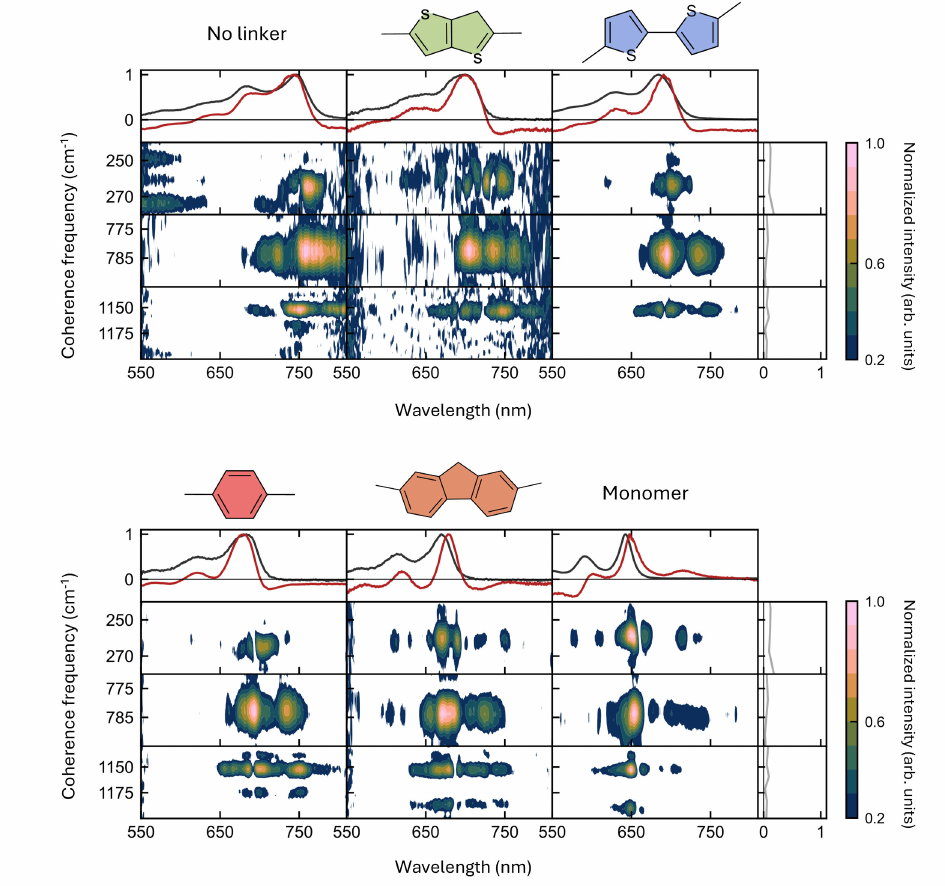} 
	\caption{ \textbf{Enhanced quantum beat amplitude on the $\mathrm{(TT_1)^1}$ photoproduct.} Impulsively excited quantum beat amplitude in the dimers as a function of probe detection wavelength. The coherence frequency $\omega_T$ is obtained by Fourier transformation (FT) of the residuals along the $T$ delay axis. The top panel overlays the absorption (solid black) and the pump-probe (red) spectrum. The pump-probe spectrum in the respective dimers corresponds to the $T$ time point where the near-IR ESA feature peaks. The delay points are 0.5~ps for $\mathrm{D_{0-2}}$, 3~ps for $\mathrm{D_{3-4}}$ and 3.8~ps for the monomer. The color levels are drawn at every 10\% interval between the range 20\% to 100\%. The right inset panel at the end shows the FT spectra of the chlorobenzene solvent. The quantum beat amplitude is increasingly enhanced in the near-IR ESA band corresponding to the $\mathrm{(TT_1)^1}$ photoproduct and correlates with the $\mathrm{(TT_1)^1}$ formation rate. }
		\label{fig:fig3}
\end{figure*}

Fig.~\ref{fig:fig3} presents quantum beat maps as a function of detection wavelength for the dimers, following an impulsive pump excitation. Details of the experiment and data analysis are presented in Supplementary Note S3.4. The beating frequencies all correspond to intramolecular vibrational modes of the Pc chromophore, consistent with the ground state Raman spectrum ({Supplementary Fig.~S26}), and no new frequencies are seen in the dimers compared to the monomer (Fig.~\ref{fig:fig3} (bottom right). Interestingly, compared to the monomer, where the quantum beat amplitude is concentrated within the main absorption and emission bands, the dimers show a distinctly different beating amplitude distribution. The beating amplitude is concentrated under the $\mathrm{(TT_1)^1}$-specific near-IR ESA band, and not under bluer wavelengths where the emissive species peaks, as seen in FLEX spectra (Fig.~\ref{fig:fig2}). The beating amplitude trends tend towards those of the monomer as the $\mathrm{(TT_1)^1}$ formation rates are progressively slower in going from $\mathrm{D_{0-4}}$. We have also checked that for very weakly coupled dimers such as those with twisted bi-phenyl linkers\cite{PatilJACS2023} where near-IR ESA band is weak/absent, the beating amplitude distribution resembles that of Pc monomer (Supplementary Fig.~S28). Additionally, we have also checked that excitation with a pump bandwidth that is limited to longer than 700~nm so as to only excite the red-most absorption band also reproduces similarly enhanced quantum beats in the $\mathrm{(TT_1)^1}$ photoproduct ({Supplementary Fig.~S25}). 

\clearpage

Role of intramolecular vibrations in promoting $\mathrm{(TT_1)^1}$ formation has also been proposed\cite{CamposACSCentSci2016, Herbert2017,TempelaarJCP2018,Duan2020,Bhattacharyya2023Low} in various theoretical works. Based on the normal mode analysis by Thoss and co-workers\cite{ThossJChemPhys2019}, many of such modes correspond to in-plane/out-of-plane deformations and ring-breathing motions of the pentacene framework. Joo\cite{Kim2020Non,Heo2025Tracking} and Petelenz\cite{Petelenz2019} have shown that enhanced quantum beats on the photoproduct report on the nuclear modes that are most displaced during ultrafast internal conversion, where enhanced beating amplitudes are suggestive of large nuclear reorganization involved during the internal conversion. Our observation of enhanced, but non-selective, quantum beat amplitudes on the $\mathrm{(TT_1)^1}$ photoproduct of strongly coupled dimers with $\mathrm{(TT_1)^1}$-specific near-IR ESA band suggests that large nuclear reorganization is involved during $\mathrm{(TT_1)^1}$ formation and its subsequent evolution away from the FC region. \par 

A natural question to ask at this point is whether the observed large nuclear reorganization upon $\mathrm{(TT_1)^1}$ formation, and the associated structural relaxation, plays any role in decorrelating the two coupled triplets in the $\mathrm{(TT_1)^1}$ state. We leverage polarization-selective impulsive pump-probe and polarization anisotropy measurements to investigate this question. We will demonstrate that an electronic character evolution associated with $\mathrm{(TT_1)^1}$ decorrelation, eventually into free triplets, can be directly inferred using these measurements.  


\subsection*{Polarization-selective pump-probe directly reports singlet electronic character in the $\mathrm{(TT_1)^1}$ state}
\begin{figure*}[ht!]
	\centering
	\includegraphics[width= 3 in ]{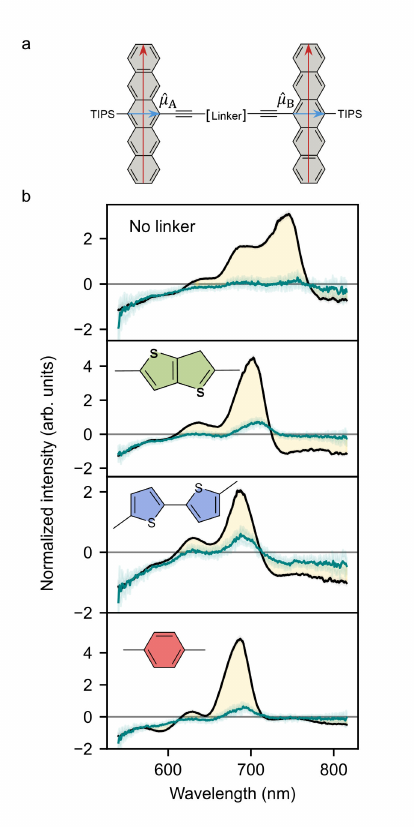} 
	\caption{\footnotesize \textbf{The $\mathrm{(TT_1)^1}$ near-IR ESA band originates from a mixture of short- and long-axis polarized transitions.}  \textbf{a} The electronic transition dipole orientation in the molecular frame. Blue arrows denote the S$_0$$\rightarrow$S$_1$ transition dipoles, denoted by $\hat{\mu}_A$ and $\hat{\mu}_B$. ${\mathrm{(TT_1)^m \rightarrow (TT_n)^1m}}$ long-axis polarized transition dipoles, which originate from the correlated triplet pair state and contribute to the near-IR ESA band, are denoted by the red arrows. \textbf{b} Magic angle (MA) versus polarization-selective pump-probe (POL-PP) spectra at $T$ = 0.5 ps for all the dimers. Both spectra are normalized to the ESA signal around 550 nm. The standard deviation resulting from the averaged PP spectra (over a 0.1~ps window) is overlaid as an error band. The suppression of the near-IR ESA band implies that the $\mathrm{(TT_1)^1}$-specific near-IR ESA band must involve both short- and long-axis polarized transitions. The relative peak ratios of different bands have been tabulated in Supplementary Table~S10.}
	\label{fig:fig4}
\end{figure*}

The S$_0$ $\rightarrow$ S$_1$ pump induced transition is short-axis polarized\cite{Shukla2007,Kohn2008,Schreiber2012} in Pc chromophores. {In comparison, the $\mathrm{(TT_1)^m}$ $\rightarrow$ $\mathrm{(TT_n)^m}$ probed transition, that leads to ESA from the correlated triplet species, is polarized along the long-axis\cite{Ginsberg2018,Mazumdar2018,Mazumdar2020,MazumdarJPCLett2017,Friend_PRB_2011,Schreiber2012,Taylor2009} of the chromophore similar to T$_1$ $\rightarrow$ T$_n$ in free triplets}. The transitions are denoted in Fig.~\ref{fig:fig4}a. The orthogonal nature of pump- and probe-induced transition dipoles implies that polarization schemes that select for orthogonally polarized transition dipoles\cite{ZanniPNAS2022,Bhattacharyya2023Low,sahu2025isolating,ThomasNatComm2026} can be applied in a pump-probe geometry as a $( 0^{\circ}0^{\circ}+60^{\circ}-60^{\circ})$ polarization sequence, as illustrated in Fig.~\ref{fig:fig1}c. Here $0^{\circ}$ denotes vertical polarization of the first two pump pulses. $+60^{\circ}$ and $-60^{\circ}$ denote the polarization of the probe and the analyzer in the transmitted probe, respectively. The sequence denotes the polarization of the electric field in each of the total four light-matter interactions of a four-wavemixing experiment. It can be shown\cite{ZanniPNAS2022,Bhattacharyya2023Low,sahu2025isolating} that a $(0^{\circ}0^{\circ}+60^{\circ}-60^{\circ})$ measurement, denoted as $\mathrm{S_{POL}}$, is equivalent to a difference signal between parallel ($\mathrm{S_{PA}}$) and perpendicularly ($\mathrm{S_{PE}}$) polarized pump and probe as $\mathrm{S_{POL} = S_{PA} - 3 S_{PE}}$. This eliminates isotropic contributions which obey $\mathrm{PA = 3PE}$. The strength of the surviving terms is proportional to $\sim$ sin$^2\theta_{\mathrm{AB}}$, where $\theta_{\mathrm{AB}}$ is the angle between pump and probe induced transition dipoles. These ideas have been described in recent works\cite{ZanniPNAS2022,Bhattacharyya2023Low,sahu2025isolating} and explained further in Supplementary Note 4. Our recent work has applied this pulse sequence to select orthogonally polarized $\mathrm{Q_x - Q_y}$ transitions in porphyrin nanotubes \cite{ThomasNatComm2026}. In the context of iSF dimers, this polarization sequence, denoted as $\mathrm{POL}$, eliminates nonlinear signals due to parallel pump and probed induced transition dipoles. The resulting signal maximizes for mutually perpendicular dipoles, such as S$_0$ $\rightarrow$ S$_1$ and $\mathrm{(TT_1)^m}$ $\rightarrow$ $\mathrm{(TT_n)^m}$. The physical rotation of the molecule on nanosecond or longer timescales is neglected in this argument. Note that the design of Pc dimers is such that the short-axis of the two Pc chromophores are always near-parallel, that is, $\theta_{\mathrm{AB}} \approx 0$, thereby enabling the application of the above idea to select for orthogonally polarized transitions arising from the triplets manifolds. \par

Fig.~\ref{fig:fig4}b compares the magic angle (MA) versus the POL-PP spectra for dimers $\mathrm{D_{0-3}}$ for which the diminished POL signals exhibited reasonable SNR.  Both spectra are normalized to the blue ESA band. {The spectra are shown for $T = $ 0.5 ps where only the $\mathrm{(TT_1)^1}$ species within the triplet manifold is expected}. The POL-PP spectra show a significant but expected suppression of the positive ground state bleach (GSB). The GSB band suppression is expected because it originates from pump and probe induced transition dipoles, S$_0$ $\rightarrow$ S$_1$, which are both parallel and polarized along the short-axis. In contrast, a pronounced suppression of the near-IR ESA band in all cases is unexpected and indicates that it cannot be described solely by the long-axis polarized $\mathrm{(TT_1)^1}$ $\rightarrow$ $\mathrm{(TT_n)^1}$ probed transition dipoles. Instead, the transitions contributing to this band must contain significant short-axis polarized electronic character. It should be emphasized that intermediate polarized nature of the near-IR ESA band, seen as its suppression under POL sequence, is found to be general across the bridging motifs suggesting a limited role of the bridge in tweaking the electronic character of the bound triplet pair from which the near-IR ESA band arises. We will come back to this point during the discussion of polarization anisotropy in Fig.~\ref{fig:fig5}. \par

Unlike other features, the blue ESA band in the POL spectra shows a prominent negative signal implying that the blue and near-IR ESA bands are significantly different in their electronic character -- the electronic transitions probed in the blue ESA band contain significantly more contribution from the near-orthogonal pumped and probed transition dipoles. This interpretation is consistent with the calculations\cite{MazumdarJPCLett2017} of Mazumdar and co-workers which suggest that the band near $\sim$550~nm originates from ${\mathrm{(S_0T_1) \rightarrow (S_0T_n)}}$ long-axis polarized ESA transitions. Note, however, that triplet absorption in the same wavelength region, arising from the heterogeneous ground state conformational landscape, can significantly complicate further interpretation of this band. \par

\clearpage


\subsection*{Minimal electronic reorientation throughout $\mathrm{(TT_1)^1}$ lifetime revealed by pump-probe polarization anisotropy}

To determine {whether the $\mathrm{(TT_1)^1}$ state decorrelates into the long-axis polarized  $\mathrm{(TT_1)^m}$ quintet or triplet species prior to molecular rotational diffusion on nanosecond timescales, we probe the electronic polarization anisotropy of the near-IR ESA band}. Polarization anisotropy tracks the electronic character evolution accompanying photophysical processes, for example, during passage through a Jahn-Teller conical intersection in square-symmetric molecules\cite{Farrow2008polarization}. The anisotropy is defined in our work as, $r(T) = (S_{\mathrm{PA}} -S_{\mathrm{MA}})/2S_{\mathrm{MA}}$, where $S_{\mathrm{PA}}$ (parallel) corresponds to a transient signal where the relative polarization between the pump and the probe is 0$^{\circ}$, whereas $S_{\mathrm{MA}}$ (magic angle) corresponds to the same with a magic angle (54.7$^{\circ}$) relative polarization. Similar definitions exist with $\mathrm{PA}$ and $\mathrm{PE}$ signal combinations.

After pumping a S$_0$ $\rightarrow$ S$_1$ short-axis polarized transition, $\mathrm{S_1 \rightarrow (TT_1)^1}$ internal conversion leads to the growth of $\mathrm{(TT_1)^1}$-specific near-IR ESA band. The POL-PP data in Fig.~\ref{fig:fig4} shows that the near-IR ESA band originating from the initially formed correlated triplet pair exhibits an intermediate polarization. Electronic structure calculations on iSF dimers suggest\cite{ Mazumdar2018,Mazumdar2020,MazumdarJPCLett2017} that ESA transitions originating from weakly correlated triplets, such as $\mathrm{(TT_1)^m}$ $\rightarrow$ (TT$_n$)$^m$ and $\mathrm{S_0T_1 \rightarrow S_0T_n}$ free-triplet-like transitions, are long-axis polarized. Similarly, experiments on Pc thin films have also revealed prominent near IR ESA band\cite{Friend_PRB_2011,Taylor2009} which lasts for tends of nanoseconds and assigned\cite{Schreiber2012,Ginsberg2018} to long-axis polarized $\mathrm{T_1 \rightarrow T_n}$ transitions. Based on these suggestions, as the initially formed triplet pair decorrelates, the polarization anisotropy of the near-IR ESA band can be expected to approach $-0.2$.\par

\begin{figure*}[!ht]
	\centering
	\includegraphics[width=1\textwidth]{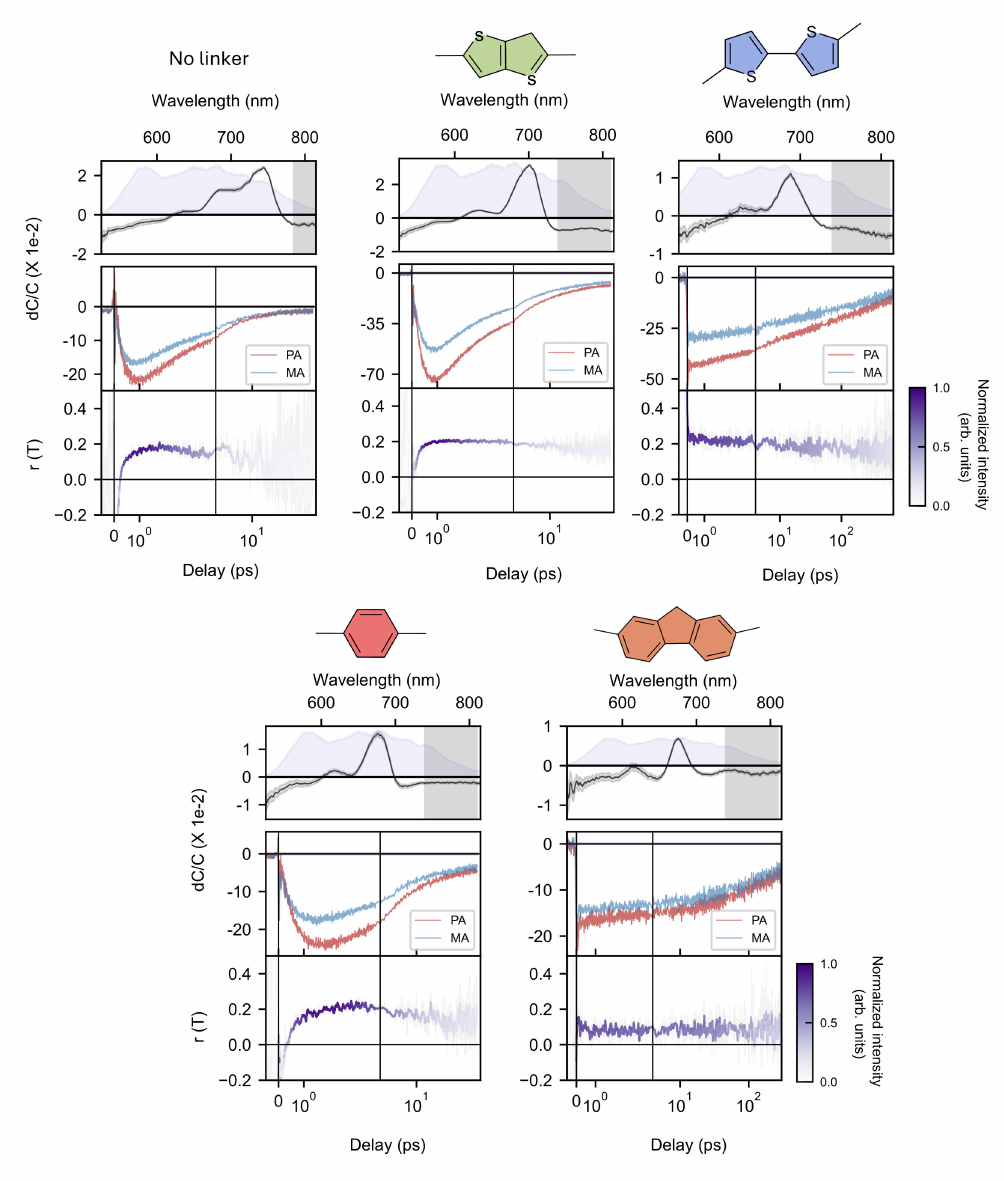}
	\caption{\footnotesize \textbf{ Pump-probe polarization anisotropy reveals flat anisotropy profile indicating minimal electronic reorientation of the $\mathrm{(TT_1)^1}$ state through its lifetime.} The top panel shows the MA pump-probe spectra corresponding to the peak amplitude in the near-IR ESA band. The delay points are 0.5~ps for $\mathrm{D_{0-2}}$ and 3~ps for $\mathrm{D_{3-4}}$. The near-IR ESA band is shown with a gray vertical band. The blue shaded trace depicts the pump spectrum. The middle panel shows PA (red) and MA (blue) pump-probe transients for the spectrally integrated band. Polarization anisotropy, calculated using the PA and MA data is shown in the bottom panel. The color bar on the anisotropy ranges from 0 (white) to 1 (blue), where 1 corresponds to the maximum of the absolute PA signal. The raw anisotropy trace is shown as a faint gray trace whereas the purple color-coded trace has been obtained by doing a 5-point running average of the raw data. The anisotropy values are tabulated in the Supplementary Table~S9. The early $T$ anisotropy is shown in Supplementary Fig.~S31. }
	\label{fig:fig5}
\end{figure*}

Fig.~\ref{fig:fig5} (top panel) displays the MA pump-probe spectra for dimers $\mathrm{D_{0-4}}$, with the near-IR ESA indicated by a gray band. The middle panel shows the PA and MA transients, spectrally integrated over a 30 (70)~nm window centered at 800 (775)~nm for $\mathrm{D_0}$ ($\mathrm{D_{1-4}}$). The bottom panel presents the corresponding polarization anisotropy. We confirmed polarization purity using Oxazine-170 in methanol ({Supplementary Fig.~S32}), yielding an early $T$ anisotropy of $\sim$0.391(3E-3) at $T\sim$0.6~ps and a late-time value of $\sim$-6.5E-3 (3E-4) at $T\sim$440~ps, approximately consistent with the theoretical limits for isolated transition dipoles undergoing rotational diffusion. \par

Interestingly, for dimers $\mathrm{D_2}$ and $\mathrm{D_4}$, the $\mathrm{(TT_1)^1}$  ESA band appears prominently within the IRF window ({Supplementary Fig.~S31}) and consistent with the corresponding early $T$ 2D spectra (Supplementary Fig.~S24). {We have also tested a 2,2'-linked version\cite{PatilAngew2024} of the $\mathrm{D_0}$ dimer (Fig.~S33) which also exhibits prompt $\mathrm{(TT_1)^1}$ formation as seen from the near IR ESA band.} Coherent vibronic mixing between the $\mathrm{S_1}$ and the $\mathrm{(TT_1)^1}$ states\cite{Bhattacharyya2023Low} in the FC region can lead to direct optical excitation of $\mathrm{(TT_1)^1}$-specific features. Similar features have been reported\cite{Musser2024,Zhu2012,Zhu2017,Turner2017} previously as well. Interestingly, we find that the prompt appearance of the pump-probe signal is also accompanied by anisotropic polarization response. \par

Although the pump-probe signals and the corresponding anisotropy during $\mathrm{(TT_1)^1}$ formation exhibit rich complexity, a detailed analysis of this early-time regime is not required for the present objectives of tracking the subsequent electronic character evolution of the $\mathrm{(TT_1)^1}$ state during triplet pair decorrelation, if any. Accordingly, we analyze the anisotropy dynamics starting from the point of maximum amplitude, corresponding to peak $\mathrm{(TT_1)^1}$ formation. {Notably, in all 6,6'-linked dimers, the anisotropy reaches a plateau between $\sim$0.1-0.2 and, contrary to the expected evolution towards $-0.2$ as the initially formed triplet pair decorrelates, remains essentially unchanged throughout the lifetime of the near-IR ESA band. } The anisotropy in case of $\mathrm{D_2}$ and $\mathrm{D_4}$ exhibits a slow decay towards zero because the triplet pair lifetime approaches the rotational diffusion time of the molecule. The anisotropy plateau is summarized in {Supplementary Table~S9}, and independently confirms the intermediate polarization of the near-IR ESA band as suggested by the POL-PP spectra in Fig.~\ref{fig:fig4}. At a later time delay, the traces become noisy as the MA signal approaches zero. To elucidate this, the $r(T)$ traces are color-coded with a colorbar that ranges from 1 (blue, maxima of the absolute PA signal) to 0 (white, minima of the absolute PA signal). We have also checked that the above lack of electronic reorientation persists even under red-edge photoexcitation, as shown in {Supplementary Fig.~S30}.  Note that the interpretation of the GSB/SE band anisotropy is complicated by signal overlap from multiple ground-state conformations. The observed traces, with initial anisotropy ranging from $\sim$0.27-0.40, are shown in \ab{Supplementary Fig.~S29}. \par



\clearpage

Similar anisotropy behavior -- negligible electronic reorientation following the formation of the $\mathrm{(TT_1)^1}$ state -- across a variety of bridging motifs has interesting implications. {Several recent studies \cite{Hasobe2020_JPCB, Johnson2020_NatChem,Tayebjee2019,McCamey2023}} have suggested that THz torsional fluctuations between thermally accessible conformations can suppress the inter-triplet exchange couplings to drive the formation of the quintet species. However, our room-temperature experiments on conformationally flexible systems suggest that, {when the triplet pair is strongly bound and leads to a prominent near-IR ESA band, THz vibrational motions, while prominent in beat maps and suggestive of large nuclear reorganization during $\mathrm{(TT_1)^1}$ evolution, are ineffective at decorrelating the $\mathrm{(TT_1)^1}$ state before its eventual decay.} Note that the case of $\mathrm{D_2}$ and $\mathrm{D_4}$, where the anisotropy slowly decays toward zero because the triplet pair lifetime approaches the rotational diffusion time of the molecule, further indicates that in the dimers studied here, the mechanism of triplet pair decorrelation through THz torsional motions becomes significant only on timescales exceeding $\sim$100 ps. Negligible electronic reorientation with anisotropy plateau at $\sim$0.1-0.2 across the bridging motifs suggest that the $\mathrm{(TT_1)^1}$ state remains strongly bound in all the 6-6'-linked dimers which show a prominent near IR ESA band. Rather than evolving into quintet or triplet spin states, which would be accompanied by long-axis polarized transitions, the majority of the $\mathrm{(TT_1)^1}$ population decays via internal conversion\cite{Scholes2018} before triplet-pair decorrelation can occur through structural fluctuations. Our above observations are consistent with the electronic structure calculations\cite{Mazumdar2020} of Mazumdar et al. -- in case of the para- linked Pc dimer, similar to $\mathrm{D_3}$, the $\mathrm{(TT_1)^1}$ state is predicted to not dissociate into free triplets through its lifetime. Interestingly, the EPR spectra of these dimers\cite{PatilJACS2023}, recorded as early as $\sim$0.8~$\mu$s, indicate a suppressed quintet population as $\mathrm{(TT_1)^1}$ decay accelerates from $\mathrm{D_4}$ to $\mathrm{D_1}$ owing to increased electronic coupling. {This trend is also consistent with our anisotropy measurements and highlights the complementary role of polarization anisotropy in tracking the electronic character evolution of triplet pair prior to longer timescale spin-specific measurements.} \par

Electronic structure calculations by Mazumdar et al.\cite{Mazumdar2018} further suggest that the intensity of the $\mathrm{(TT_1)^1}$ near-IR ESA band and its tendency to decorrelate into free triplets is highly sensitive to the bridge and linkage geometry, which modulate electron and hole overlap integrals through changes in intermolecular separation and torsional distortion. For example, Fig.~5 of ref.\cite{Mazumdar2018} shows that, with weaker electronic coupling, ESA bands originating from the $\mathrm{(TT_1)^1}$ state start resembling those originating from free triplet absorption in the dimer with triplets localized on individual Pc units. {These ideas are further extended in recent\cite{Campos_JACS2026_metaBP1} experiments from Campos, Sfeir and coworkers where the bridge linkage is also found to dictate the inter-chromophore spin dipolar interactions. In our experiments on a 2,2$'$-linked analogue\cite{PatilAngew2024} of the $\mathrm{D_0}$ dimer (Fig.~S33), where the near-IR ESA band is substantially weaker than in the corresponding 6,6$'$-linked dimer ($\mathrm{D_0}$ in Fig.~\ref{fig:fig5}), the anisotropy of the near-IR ESA band already starts at $-0.2$. The anisotropy subsequently decays slowly toward zero as the $\mathrm{(TT_1)^1}$ lifetime exceeds the rotational diffusion time of the molecule, analogous to the behavior observed for the 6,6$'$-linked $\mathrm{D_2}$ and $\mathrm{D_4}$ dimers in Fig.~\ref{fig:fig5}, although from a markedly different initial anisotropy. Thus, the electronic character of the $\mathrm{(TT_1)^1}$ state in the 2,2$'$-linked dimer is fundamentally distinct from that of its 6,6$'$-linked analogue ($\mathrm{D_0}$). The triplet pair in the former does not contain electronic admixtures of singlet character, is weakly bound to begin with, and exhibits predominantly long-axis polarized transitions closely resembling the decorrelated triplet-pair species expected at longer times. In this sense, the $\mathrm{(TT_1)^1}$ state is effectively primed for decorrelation through structural fluctuations on longer timescales. Consistent with this picture, the 2,2$'$-linked dimer, 2Ac-P2 in ref.\cite{PatilAngew2024}, yields 2--3 times more free triplets than its 6,6$'$-linked counterpart, 6Ac-P2 in ref.\cite{PatilAngew2024} and $\mathrm{D_0}$ in this study, although with an overall low triplet yield. The short $\mathrm{(TT_1)^1}$ lifetime of 0.86~ns, however, precludes conventional EPR measurements. The above contrast again demonstrates that polarization-selective pump--probe and anisotropy measurements provide a direct optical probe of triplet pair decorrelation, complementing spin-specific techniques that monitor its spin evolution at longer timescales.}



Next we investigate the question of why the near-IR ESA band shows intermediate polarization. To address this question, we performed electronic structure calculations on the strongly coupled $\mathrm{D_0}$ dimer using the screened configuration interaction singles and doubles (scrCISD) method\cite{Jain2025} recently developed by one of us. In this method, an effective many-body Hamiltonian that incorporates screening is constructed and projected onto the combined space of single and double excitations. Diagonalization of this Hamiltonian yields the adiabatic electronic eigenstates and corresponding wavefunctions. Unlike methods based on the Pariser--Parr--Pople (PPP) Hamiltonian\cite{MazumdarJPCLett2017}, scrCISD is \textit{ab initio} and does not rely on empirical parameters. The inclusion of both single- and double-excitation manifolds enables an accurate description of singlet--triplet mixing in the adiabatic states. \par 

\begin{figure*}[!ht]
	\centering
	\includegraphics[width=5in]{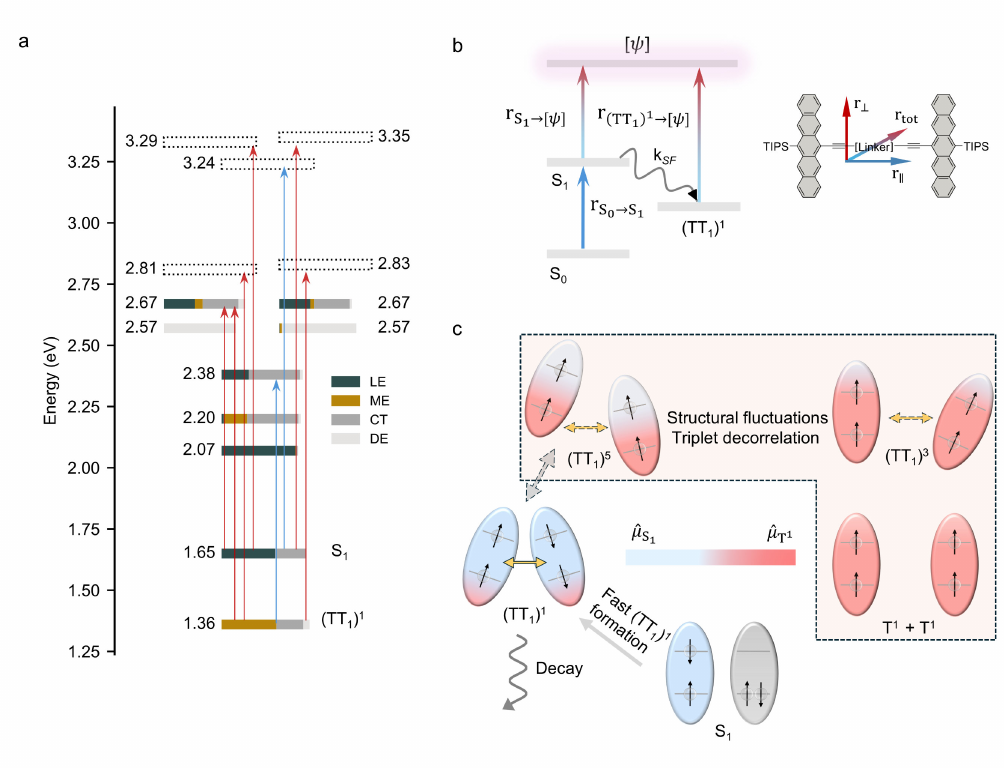}
	\caption{\footnotesize \textbf{scrCISD calculations are consistent with intermediate polarization of the near-IR ESA band.} \textbf{a} Electronic character of adiabatic electronic states at the FC geometry for the $\mathrm{D_0}$ dimer in a planar configuration, calculated using the scrCISD method\cite{Jain2025}. The character of each state is decomposed into \mj{localized basis states as} represented by the color-coded segments in the panel. \mj{The uncharacterized states are denoted with dashed black bars.} Blue and red arrows mark the dominant transition dipole moments with color denoting the dominant polarization direction along short and long axis, respectively. The transitions are shown for excitations originating from the $\mathrm{S_1}$ and $\mathrm{(TT_1)^1}$ states in the near-IR and shortwave IR regions. Supplementary Tables~S11-S12 show all the predicted transitions. \textbf{b} Simplified energy level schematic of the electronic manifold. The wavy arrow denotes the singlet fission process from the initially excited $\mathrm{S_1}$ to the $\mathrm{(TT_1)^1}$ state. Even though individual transitions to the higher-lying electronic states, denoted by $[\psi]$ manifold, are dominantly only $\mathrm{\hat{x}}$ or only $\mathrm{\hat{y}}$ polarized, the net transition strength and anisotropy will result in intermediate polarization. {This is denoted by mixed blue and red transitions}. In the inset, the blue and red vectors correspond to the theoretical anisotropy limits for parallel ($r_{\mathrm{\parallel}}=0.4$) and mutually orthogonal ($r_{\mathrm{\perp}}=-0.2$) pumped and probed transition dipoles in the molecular frame. The mixed red-blue vector represents the effective transition dipole moment contributing to the net anisotropy of the near-IR ESA band. \textbf{c} A schematic illustrating the expected evolution of electronic transition density during iSF versus that observed for the Pc dimers with near-IR $\mathrm{(TT_1)^1}$-specific feature. Light blue and red shade denote short axis polarized singlet and long axis polarized free triplet transitions. Ideally, structural fluctuations following $\mathrm{(TT_1)^1}$ formation should eventually lead to the formation of uncorrelated triplets ($\mathrm{T^1} + \mathrm{T^1}$) via $\mathrm{(TT_1)^5}$ quintet and $\mathrm{(TT_1)^3}$ triplet manifolds. Experimentally observed intermediate polarization and flat anisotropy profiles in the $\mathrm{(TT_1)^1}$ near-IR band suggest that $\mathrm{(TT_1)^1}$ dominantly decays via internal conversion before any triplet decorrelation, and maintains mixed singlet-triplet electronic character, as a $\mathrm{[S_1 + (TT_1)^1]}$ species, throughout its lifetime. The broken gray arrow, dashed box indicate the negligible formation of the $\mathrm{(TT_1)^5}$ and subsequent states.}
	\label{fig:fig6}
\end{figure*}

The resulting adiabatic states were further projected onto a localized orbital basis obtained via Boys localization, allowing their decomposition into locally excited (LE), CT, multiexcitonic (ME), and doubly excited (DE) contributions at the FC geometry, as defined in ref.~\cite{GuldiJPCLett2022} \mj{and in Supplementary Note 5}. Transition dipole moments originating from S$_1$ (predominantly LE) and from $\mathrm{(TT_1)^1}$ (predominantly ME) to higher-lying excited states can be resolved into components along the short ($\mathrm{\hat{x}}$) and long ($\mathrm{\hat{y}}$) molecular axes of the dimer. In the same fashion, oscillator strength of each probed transition ending at the higher lying $\mathrm{j^{th}}$ state, can be decomposed into short ($\mathrm{\hat{x}}$) and long ($\mathrm{\hat{y}}$) axis components, and denoted as $|\mu_\mathrm{{j,x}}|^2$ and $\mathrm{|\mu_{j,y}|^2}$, respectively. The calculated states and energies are summarized in Supplementary Tables~S11-12. The tables also show all the transitions in the near-IR and shortwave IR regions which start from $\mathrm{S_1}$or $\mathrm{(TT_1)^1}$, and end at the $\mathrm{j^{th}}$ higher lying electronic state. The dominant transitions carry oscillator strength primarily along either the short or long molecular axis, with the orthogonal component being negligible. Fig.~\ref{fig:fig6}a highlights the prominent oscillator strengths among the ESA transitions starting from S$_1$ and $\mathrm{(TT_1)^1}$ states. \mj{The calculations further predict several additional states with significant CT character, giving rise to transitions spanning the near-IR and shortwave IR regions, in agreement with prior studies by Mazumdar and co-workers\cite{MazumdarJPCLett2017,Mazumdar2018,Mazumdar2020}. The state characters in the diabatic basis are tabulated in Supplementary Table~S13 are only shown up to energies of $\sim$2.67 eV. The characterization of the higher energy states would require using a significantly larger localized orbital basis. However, Boys localization for extended virtual spaces is known to exhibit slow convergence and may produce delocalized or chemically less meaningful localized orbitals \cite{Jorgensen2011}.} \par 



The total polarization anisotropy expected in the near-IR ESA band due to transitions to the higher-lying electronic states can be defined\cite{Qian2003Role} as $\mathrm{r_{tot} = \sum\limits_{j} S_j r_j/\sum\limits_{j} S_j}$, where $\mathrm{r_j}$ is the calculated anisotropy for transitions ending at the $\mathrm{j^{th}}$ state in Supplementary Tables~S11-S12. $\mathrm{S_j}$ is the isotropic signal strength which is proportional to the corresponding oscillator strength.  The corresponding anisotropy along the short and long axis is given by $\mathrm{r_{\parallel} = 0.4}$ and $\mathrm{r_{\perp} = -0.2}$, respectively, where it is implicit that the pumped transition dipole, $\hat{\mu}_\mathrm{{S_0 \rightarrow S_1}}$ is parallel to the short axis. The net anisotropy $\mathrm{r_j}$ for the $\mathrm{j^{th}}$ probed transition is then given by $\mathrm{r_j} = (|\mu_\mathrm{{j,x}}|^2 \mathrm{r_{\parallel}} + |\mu_\mathrm{{j,y}}|^2\mathrm{r_{\perp}})/|\mu_\mathrm{{j}}|^2$. This procedure is illustrated in Fig.~\ref{fig:fig6}b. The anisotropies calculated from considering transitions originating from solely $\mathrm{(TT_1)^1}$ and S$_1$, or combining both states for different spectral ranges are tabulated in {Table~\ref{tab:anisotropy_values}}. Note that in the above calculation, the time-dependent conversion of S$_1$ $\rightarrow$ $\mathrm{(TT_1)^1}$ state is not explicitly taken into account. Instead the anisotropy estimated from $\mathrm{S_1 \rightarrow [\psi]}$ or $\mathrm{(TT_1)^1 \rightarrow [\psi]}$ probed transitions, and both combined, provides bounds based on which the intermediate polarization of the near-IR ESA band can be understood. 

\begin{table}[ht]
	\centering
	\caption{Calculated anisotropy values for transitions starting from $\mathrm{(TT_1)^1}$ and S$_1$ states across different spectral ranges. The calculation corresponds to the schematic shown in Fig.~\ref{fig:fig6}a-c. The conversion of S$_1$ to $\mathrm{(TT_1)^1}$ state is not accounted for in the calculation. Instead only $\mathrm{S_0 \rightarrow S_1}$ pumped transition, and $\mathrm{S_1 \rightarrow [\psi]}$ or $\mathrm{(TT_1)^1 \rightarrow [\psi]}$ probed transitions have been included. The transition wavelengths are tabulated in Supplementary Tables~S11-S12. 
	}
	\label{tab:anisotropy_values}
	\begin{tabular}{lccc}
		\hline
		\textbf{Transition Wavelength} & \textbf{$\mathrm{(TT_1)^1}$ anisotropy} & \textbf{S$_1$ anisotropy} & \textbf{$\mathrm{r_{\text{total}}}$} \\ \hline
	
		$\mathrm{ < 2000 nm}$  & $0.215$ & $0.074$ & $0.16$ \\
		$\mathrm{ < 1220 nm}$  & $0.215$ & $0.041$ & $0.151$ \\
		$\mathrm{ < 1100 nm}$ & $-0.185$ & $0.08$ & $-0.022$ \\ \hline
	\end{tabular}
\end{table}

The anisotropy estimation in Table~\ref{tab:anisotropy_values} only including transitions in the near-IR ($<$1100 nm) is not sufficient to explain the experimental observations. The $\mathrm{(TT_1)^1}$ states $>$1000 nm, although not probed in our experiment, can exhibit distinct vibronic progressions\cite{XYZ2017} which extend to bluer wavelengths. When transition energies with up to $\sim$1200 nm wavelength are included, prominent oscillator strengths in that region predict a $\mathrm{(TT_1)^1}$ band anisotropy of 0.215 when transitions starting from only the $\mathrm{(TT_1)^1}$ state are considered, that is, S$_1$ $\to$ $\mathrm{(TT_1)^1}$ formation is complete. This is larger than the near-IR ESA band anisotropy of $\sim$0.1-0.2 observed in the experiments for all dimers (Supplementary Table~S9). At the same time, because the S$_1$ anisotropy by itself is less than 0.04, when the calculation of $\mathrm{r_{tot}}$ includes both S$_1$ and $\mathrm{(TT_1)^1}$ starting states, the resultant anisotropy, weighted by isotropic signal strengths, becomes $\sim$0.15, well within the observed range of near-IR ESA band anisotropies (Supplementary Table~S9). Note that we have also checked the consistency of the observed anisotropy values under excitation of only the red-edge of the absorption spectrum (Supplementary Fig.~S30), which preferentially excites the non-emissive planar conformations that lead to fast $\mathrm{(TT_1)^1}$ formation (Fig.~\ref{fig:fig2}). Note also that inclusion of more states in the shortwave IR, even though not probed in the experiment, does not affect this conclusion. On the other hand, neglecting transition wavelengths longer than 1100 nm would require significantly more singlet-triplet electronic mixing and oscillator strength in the $\mathrm{S_1 \to [\psi]}$ transitions to explain the observed anisotropy. From the anisotropy analysis above, we infer that the $\mathrm{(TT_1)^1}$ state, identified through the near-IR ESA band, possesses significant admixture with $\mathrm{S_1}$ in the FC region, forming a $\mathrm{[S_1 + (TT_1)^1]}$ mixed electronic state. The flat anisotropy profiles, which plateau at an intermediate polarization, further indicate that this mixed electronic character is persistent during nuclear evolution away from the FC region.

We note that while our calculations show qualitative agreement with experiments and reproduce the observed intermediate polarization of the $\mathrm{(TT)^{1}}$-specific near-IR ESA band, we emphasize trends rather than quantitative agreement, as several physical effects are not included. In particular, transition dipole moments are evaluated at the ground-state-optimized geometry and include only the electronic contributions. Vibronic effects, arising from the inclusion of vibrational modes of the initial and final electronic states, are neglected although they are likely to play an important role\cite{Bhattacharyya2023Low} in $\mathrm{S_1-(TT_1)^1}$ vibronic mixing. Additionally, the scrCISD method underestimates the energies of predominantly singly excited states due to coupling with double excitations, with the error increasing with the coupling strength. 

\subsection*{Discussion}

Taken together, the observations of ultrafast $\mathrm{(TT_1)^1}$ formation, even its direct optical excitation within the instrument-response (Supplementary Figs.~S24, S31) in some cases, its fast internal conversion to the ground state on a few picosecond timescale, intermediate electronic anisotropy of the $\mathrm{(TT_1)^1}$-specific near-IR ESA band with negligible electronic reorientation, all consistently indicate that the near-IR ESA band in strongly coupled Pc dimers is indicative of a $\mathrm{(TT_1)^1}$ state that is electronically mixed with the S$_1$ state in the FC region as a $\mathrm{[S_1 + (TT_1)^1]}$ species, with the mixed electronic character persistent throughout its lifetime. When the triplet pair is strongly bound, significant nuclear reorganization accompanying $\mathrm{(TT_1)^1}$ evolution, manifested as enhanced THz vibrational quantum beats in the photoproduct, is evidently ineffective in disrupting the undesirable singlet-triplet electronic mixing and driving triplet pair decorrelation. The case of dimers $\mathrm{D_2}$ and $\mathrm{D_4}$ is particularly interesting because the triplet pair lifetime outlasts the rotational diffusion timescale and yet the near IR ESA band exhibits intermediate polarization anisotropy which slowly decays towards zero due to physical rotation of the dimer. {The lack of long-axis polarized transitions in the near IR ESA band, expected\cite{Mazumdar2018,MazumdarJPCLett2017,Mazumdar2020} from a decorrelated triplet pair,  suggests that the mechanism of triplet pair decorrelation through THz vibrational motions is evidently slower than rotational diffusion timescales. In such a scenario, singlet electronic admixture within the $\mathrm{(TT_1)^1}$ state becomes particularly detrimental because it promotes decay via internal conversion, thereby easily outcompeting triplet-pair decorrelation through nanosecond or longer timescale structural fluctuations.} The picture that emerges from our observations is summarized in Fig.~\ref{fig:fig6}c. \par

{The contrast of 6,6'-linked dimers with a 2,2'-linked analogue is interesting with regard to the role of bridge linkage in dictating the electronic character of the initially formed $\mathrm{(TT_1)^1}$ state. For the 2,2'-linked dimer, the initial polarization anisotropy of the near-IR ESA band already indicates long-axis polarized transitions suggestive of a $\mathrm{(TT_1)^1}$ species with weakly bound triplets. Thus, complementary to recent EPR measurements\cite{Campos_JACS2026_metaBP1} which report a vital role of the bridge linkage in dictating spin dipolar interactions, polarization anisotropy directly reports a distinct electronic character of the initially formed $\mathrm{(TT_1)^1}$ species in the 2,2'-linked dimer. The initially formed $\mathrm{(TT_1)^1}$ state is already primed for triplet pair decorrelation through longer timescale structural fluctuations.} 

The polarization-selective experiments on the $\mathrm{(TT_1)^1}$ state introduced in this work confirm the prediction\cite{Mazumdar2020} by Mazumdar et al. that triplet excitons in the $\mathrm{(TT_1)^1}$ state of a para-linked Pc dimer, similar to $\mathrm{D_3}$, do not dominantly separate into free triplets. Our recent study\cite{PatraJCP2026} of a strongly coupled  naphthalenediimide dimer, where no unique $\mathrm{(TT_1)^1}$ spectral marker is available, also points to the same. The evidence for mixed singlet-triplet electronic character, reported by the intermediate polarization and anisotropy of the near-IR ESA band, is also consistent with prior observations\cite{XYZ2017,Scholes2018} that the $\mathrm{(TT_1)^1}$ species exhibits isolated chromophore-like singlet features in the visible pump-probe spectra. Our observations are in line with the hypothesis\cite{Scholes2018} of Pensack et al. that the $\mathrm{(TT_1)^1}$ state may be stabilized through configuration mixing with locally excited singlets. Furthermore, our direct optical evidence for the electronic character of the $\mathrm{(TT_1)^1}$ species is consistent with the chemical reactivity observations of Trinh \textit{et al.}\cite{XYZ2017}, wherein the $\mathrm{(TT_1)^1}$ state, identified via the near-IR ESA band, exhibits reactivity distinct from that of free triplets. 

Spectral similarity between $\mathrm{(TT_1)^1}$, $\mathrm{(TT_1)^3}$, $\mathrm{(TT_1)^5}$, and free triplets is well established\cite{GuldiWasielewski2017_NatComm_UnifiedSF}, rendering species identification challenging. {Our results identify the polarization anisotropy of the $\mathrm{(TT_1)^1}$ ESA band as a direct optical probe of the electronic character and decorrelation of strongly bound triplet pairs, complementing spin-specific probes that monitor the resulting spin dynamics at longer timescales.} Experiments\cite{Schreiber2012,Taylor2009,Friend_PRB_2011} on Pc thin films, with high yield of free triplets,  consistently show a strong near IR ESA band. Yet it lasts for tens of nanoseconds and is long-axis polarized\cite{Ginsberg2018}. Our observations suggest that molecular design strategies for iSF should either minimize the formation of strongly bound triplet pairs, identified through the anisotropy of the $\mathrm{(TT_1)^1}$-specific near-IR ESA feature, or instead promote rapid triplet diffusion via energy migration\cite{Scholes2015,Rao2024_JACS,Kim2024_JACS, Johnson2020_NatChem,Kobori2018} to neighboring chromophores, thereby increasing opportunities for triplet pair decorrelation. The efficacy of such decorrelation can be directly probed through polarization-selective pump-probe and anisotropy measurements.

 \section*{Methods}

\subsection*{Sample Preparation}
The samples were synthesized using procedures reported previously\cite{PatilJACS2023,PatilAngew2024}. Sample solutions were prepared by dissolving pentacene dimers in nitrogen-purged chlorobenzene (PhCl). To ensure complete dissolution and removal of particulate matter, solutions were sonicated for 20 minutes at $15^{\circ}\text{C}$ and filtered through $0.22\ \mu\text{m}$ PVDF membranes. Final solutions were transferred to $200\ \mu\text{m}$ pathlength quartz cuvettes. All processing steps were conducted under a nitrogen atmosphere to minimize oxygen exposure. Optical densities were maintained between 0.3 and 0.45, with steady-state absorption monitored before and after experiments to confirm the absence of photodegradation or aggregation.

\subsection*{Optical Setup}
The optical setup is a partially collinear white-light two-dimensional electronic spectroscopy (2DES) and pump-probe (PP) setup reported in our previous works\cite{Thomas2023Rapid, Bhat2023Rapid}. White-light continua (WLC) were generated by focusing 100~kHz Yb:KGW generated 1040 nm pulses -- $\sim$300fs, $\sim$1 $\mu$J -- in YAG crystals (4~mm for pump; 8~mm for probe). The optical dispersion was pre-compensated using a combination of glass wedges and chirped mirror pairs to maximize the two-photon interferometric autocorrelation recorded using a SiC photodiode. This yielded $\sim$10-15~fs pulse duration for the pump pulse (Supplementary Figure~1). This pulse duration was found to be sufficient to impulsively excite vibrational wavepackets up to $\sim$1450~cm$^{-1}$. Global fitting of the spectrally resolved pump-probe data across a $\sim$300~nm bandwidth indicated a typical instrument response function (IRF) full width at half maximum (FWHM) of 30--40~fs across all samples.

Establishing polarization purity was critical for the polarization-selective pump-probe and anisotropy studies. The setup incorporated achromatic half-waveplates (Thorlabs, AHWP05M-600) and thin-film linear polarizers (Thorlabs, LPVISC-050-MP2) to maintain independent control of pump and probe polarizations. At the sample location, we achieved extinction ratios of 993 for the pump and 1428 for the probe arms.  The above polarization settings were validated for a $\mathrm{D_{2h}}$ symmetric phthalocyanine molecule as detailed in our recent work\cite{ThomasNatComm2026}. We further validated the polarization configuration through anisotropy measurements on Oxazine-170 reference (Supplementary Figure~32). Further details of the setup are mentioned in Supplementary Note 1.1.

\subsection*{Data Collection Scheme}
The pump-probe and 2DES measurements are based on the rapid scan methods established\cite{Bhat2023Rapid,Thomas2023Rapid} in our previous works. For pump-probe experiments, $T$ was scanned from -0.5~ps to 660~ps across four intervals with constant scan velocities tabulated in Supplementary Table~S4. Anisotropy data were collected by recording interleaved PA and MA scans so that different polarization errors are encountered in each of the four PA or MA trials. This ensures statistical robustness because polarization errors can be averaged out.\par 

In 2DES kinetic rate scans, the coherence time ($\tau$) was scanned continuously using a common-path interferometer (CPI) with $\sim$15-20 $T$ time steps per decade. A slow $\tau$ scan velocity (0.4 mm/s) was employed to finely sample $\tau$. This allows for effective suppression of pump-scatter through averaging multiple $\tau$ points into a binned time step (0.1346~fs). The population time points were scanned in a randomized order to mitigate systematic drift. For 2DES coherence scans, $\tau$ was scanned stepwise from -~5~fs to 75~fs with a timestep of 0.4038~fs, while $T$ stage was scanned continuously from -0.25~ps to 0.96~ps with total 1210 $T$ points per binned $T$ time step of 5~fs. With only the allowed 2N zero-padding the $\omega_{\tau}$ frequency resolution is 0.043 rad/fs. These settings are summarized in Supplementary Note~2. \par 

Polarization-selective pump-probe (POL-PP) scans were also performed with a $(0^{\circ}0^{\circ}+60^{\circ}-60^{\circ})$ configuration, recording 30 continuous $T$ scans per dimer with a binned $T$ time step of 10~fs.
Further details are mentioned in Supplementary Note~2. \par

\subsection*{Data Processing}
All transient datasets were chirp-corrected using a first-order polynomial derived from global fitting to ensure that the temporal zero remained independent of detection wavelength. Even without any chirp correction, the risetime dispersion across $\sim$300 nm bandwidth is found to be less than 39 fs (Supplementary Figure~1). For kinetic RM analysis, 2DES data cubes were globally fitted with a tri-exponential model convolved with a Gaussian instrument response. For global 2DES fitting, decay time constants were obtained from global fitting of spectrally integrated 2DES datasets and kept fixed. Further details are mentioned in Supplementary Note~3. 

The raw anisotropy is calculated from interleaved PA, MA scans for multiple trials. A running average is performed on the raw anisotropy data to obtain the color-coded purple trace shown in Fig~\ref{fig:fig5}. The averaging time window employed is over 5 data points i.e. 50~fs until 3.8~ps, 500~fs until 46~ps and 5~ps until the end of the scan. The data shown is closest to the mean value of the trial-averaged anisotropy (see Supplementary Table~9 and Supplementary Note~S3.5). 

Vibrational quantum beat maps were generated by isolating oscillatory components from pump-probe residuals. Following the subtraction of exponentially decaying backgrounds using fourth degree polynomials, the residuals were zero-padded to $\mathrm{4N}$ points and windowed with a hyperbolic tangent filter prior to Fourier transformation. With only the allowed 2N zero-padding, the $\omega_{T}$ frequency resolution is 4.6 cm$^{-1}$. Further details are mentioned in Supplementary Note 3.

\subsection*{\mj{Electronic Structure Calculations}}
The electronic structure calculations were performed using the screened configuration singles and doubles method\cite{Jain2025} (scrCISD) on a planar $\mathrm{D_0}$ dimer. This allowed us to estimate adiabatic electronic characters at the FC geometry. States were projected into a localized orbital basis (Boys localization) to resolve LE, CT, ME and DE characters. The transition dipole moments from S$_1$ (dominantly LE) and $\mathrm{(TT_1)^1}$ (dominantly ME) were decomposed into components along the short-axis ($\mathrm{\hat{x}}$) and long-axis ($\mathrm{\hat{y}}$) of the dimer frame. Further details are mentioned in Supplementary Note 5.

 \section*{Data Availability}
All data required for interpretation and verification of experimental results are given in the paper and the Supplementary Information. 


\section*{Acknowledgements}
{\footnotesize A.B. acknowledges the Prime Ministers' Research Fellowship, MoE, India. S.P. and N.V. acknowledge the research fellowship from IISc Bangalore. M.J. and N.V. gratefully acknowledge the Nano Mission of the Department of Science and Technology, India for financial support under Grant No. DST/NM/TUE/QM-10/2019.}

\section*{Author Contributions}
{\footnotesize V.T. conceived and supervised the project. A.B. and S.P. performed the optical measurements. A.B. developed the data analysis pipeline, carried out the data analysis, and prepared all figures. N.V. performed the electronic structure calculations and wrote the corresponding sections of the manuscript and Supplementary Information under the supervision of M.J. K.M. and V.H. synthesized the samples. V.T. wrote the manuscript with contributions from A.B. and N.V. All authors discussed the results and commented on the manuscript.}

\section*{Competing Interests}
{\footnotesize The authors declare no competing interests.}

\end{document}